\documentclass{PoS}

\def\ari          {ARIANNA}
\def\hra          {HRA}
\def\myname       {Corey Reed}

\newcommand {\fig}[1]{Fig.~\ref{#1}}

\newcommand {\Fig}[1]{Figure~\ref{#1}}

\newcommand {\sect}[1]{Sect.~\ref{#1}}

\newcommand {\dg}    {\mbox{$^\circ$}}

\newcommand {\gev}   {\mbox{${\rm GeV}$}}

\newcommand {\mom}   {\mbox{\rm GeV$\kern-0.15em /\kern-0.12em c$}}
\newcommand {\mmom}  {\mbox{\rm MeV$\kern-0.15em /\kern-0.12em c$}}
\newcommand {\mass}  {\mbox{\rm GeV$\kern-0.15em /\kern-0.12em c^2$}}
\newcommand {\mmass} {\mbox{\rm MeV$\kern-0.15em /\kern-0.12em c^2$}}

\newcommand {\km}    {\mbox{${\rm km}$}}
\newcommand {\m}     {\mbox{${\rm m}$}}

\newcommand {\ns}    {\mbox{${\rm ns}$}}

\newcommand {\mhz}   {\mbox{${\rm MHz}$}}
\newcommand {\ghz}   {\mbox{${\rm GHz}$}}

\title{Performance of the {\ari} Hexagonal Radio Array}

\ShortTitle{Performance of the {\ari} Hexagonal Radio Array}

\author{\speaker{{\myname}} for the {\ari} Collaboration\\
        Department of Physics and Astronomy, University of California, Irvine\\
        Irvine, CA 92697-4575, USA\\
        E-mail: \email{cjreed@uci.edu}}

\abstract{Installation of the {\ari} Hexagonal Radio Array ({\hra}) on
  the Ross Ice Shelf of Antarctica has been completed. This detector
  serves as a pilot program to the {\ari} neutrino telescope, which
  aims to measure the diffuse flux of very high energy neutrinos by
  observing the radio pulse generated by neutrino-induced charged
  particle showers in the ice. All {\hra} stations ran reliably and
  took data during the entire 2014-2015 austral summer season. A new
  radio signal direction reconstruction procedure is described, and is
  observed to have a resolution better than a degree. The
  reconstruction is used in a preliminary search for potential
  neutrino candidate events in the data from one of the newly
  installed detector stations. Three cuts are used to separate radio
  backgrounds from neutrino signals. The cuts are found to filter out
  all data recorded by the station during the season while preserving
  85.4\% of simulated neutrino events that trigger the station. This
  efficiency is similar to that found in analyses of previous {\hra}
  data taking seasons.}

\FullConference{The 34th International Cosmic Ray Conference,\\
		30 July- 6 August, 2015\\
		The Hague, The Netherlands}

\begin{document}

\section{The {\ari} Hexagonal Radio Array}
\label{hra}

The {\ari} Collaboration plans to construct a neutrino telescope
capable of measuring the diffuse flux of high energy neutrinos in the
$10^{8}-10^{10}~{\gev}$ range~\cite{Barwick:2014pca}. The {\ari}
Hexagonal Radio Array ({\hra}) serves as a pilot program to the full
{\ari} telescope. During the 2014-2015 austral summer, the {\hra}
installation was completed at the {\ari} detector site on the Ross Ice
Shelf of Antarctica~\cite{annaProc}. It consists of eight independent
detector stations, each of which has four Log-Periodic Dipole Antenna
(LPDA), an autonomous data acquisition (DAQ) system (with local data
storage and remote data transfer capabilities) as well as local solar
power. The stations measure radio pulses in the $50~{\mhz}-1~{\ghz}$
frequency range, making them sensitive to the radio emission produced
by charged particle showers generated by neutrino interactions in the
ice, known as the Askaryan effect~\cite{Askaryan:1962, Askaryan:1965,
  Gorham:2006fy}. Details on the station hardware may be found in
Refs.~\cite{annaProc, AriannaNIMPaper}.

\begin{figure}
   \begin{center}
     \includegraphics[width=0.9\linewidth]{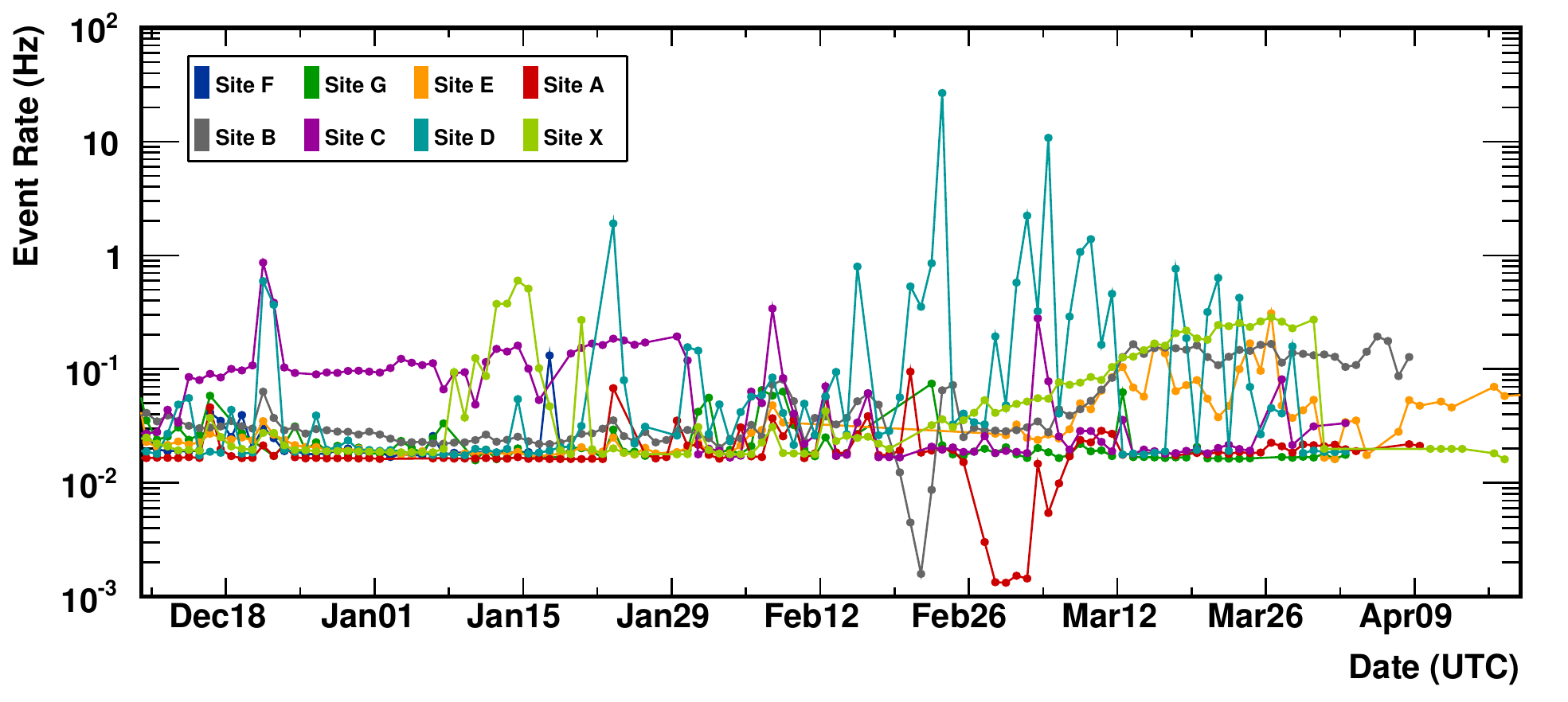}
   \end{center}
   \caption{\label{hra:fig:rates}
     Daily average trigger rates of each {\hra} station during the
     2014-2015 data taking season. See text for details.}
\end{figure}

All eight stations were installed by early December, 2014 and ran
until the sun began to set around early April, 2015. The layout of the
detector stations can be found in Refs.~\cite{Barwick:2014pca,
  annaProc}. \Fig{hra:fig:rates} shows the trigger rate of each
station during the 2014-2015 austral summer data taking season.

Similar to previous seasons, trigger rates were generally low and
minimal threshold tuning was required. While each station is similar,
small variations can lead to noticeable differences in the trigger
rates. The majority of Site~F data has not yet been transferred off
station and does not appear in the figure. During late February and
early March, the trigger requirements at Sites~A and B were
temporarily increased to lower the rates. This was done to test the
reliability of sending radio waveform data off Antarctica using
satellite communications~\cite{annaProc} limited to 340~bytes per
message. The strong solar burst on December 20, 2014 is visible in a
correlated rise of trigger rates on all stations~\cite{annaProc} (it
is also visible in \fig{ana:fig:xiVsT}, discussed in
\sect{ana:xi}). The higher rates of Site~C early in the season are
simply due to low triggering thresholds (which were later
corrected). The high rate spikes of Site~D later in the season are
caused by radio bursts emitted when the station's battery switches on
(station powered by battery) and off (station powered directly by
solar panels). This will be corrected during the 2015-2016 deployment
season by installing a new battery inside the radio-tight DAQ
box. Other rate increases are correlated with storms (i.e. February 8)
and the dramatic temperature drop that occurs during the beginning of
March. As the temperature falls from about $0{\dg}$~C to $-20{\dg}$~C,
the gain of the amplifiers increases, leading to a rise in trigger
rates unless thresholds are adjusted to compensate.

\section{Radio Signal Direction Determination}
\label{reco}

The direction from which radio signals arrive at a detector station is
reconstructed using the cross-correlation of the time dependent radio
signals observed by parallel LPDAs. Each {\hra} station has four
LPDAs, arranged symmetrically around the station DAQ box.  Antennas
situated across from each other (not adjacent) are oriented with
parallel tines and separated by 6~{\m}. Thus there are two such
parallel LPDA pairs, so that a signal perfectly co-polarized with one
antenna pair will be perfectly cross-polarized with the other antenna
pair. Cross-correlations are used to determine the likelihood that the
measured data is consistent with the expected time delays between
parallel channels for a radio signal (with a planar wave-front)
arriving from a particular direction.

The most likely radio arrival direction is determined by comparing two
separate fit procedures. The first procedure uses a genetic
minimization algorithm to choose the most likely arrival
direction. The second procedure simply scans the entire angular phase
space in $1{\dg}\times1{\dg}$ bins. In both procedures, the arrival
direction estimate is further refined using the Migrad minimization
algorithm. The two resulting fits are compared and the one with the
best likelihood is chosen.

Special calibration data has been collected in order to quantify the
radio pulse arrival direction resolution. A Pockels Cell Driver (PCD)
is used to generate a very short (about 3~{\ns}) uni-polar pulse that
is transmitted from an LPDA. The transmitting antenna is oriented with
its boresight vertically downward and placed at various positions on
the ice around a station. The calibration pulse travels downward
through the firn and ice, reflects off the sea water below the ice
shelf and then travels up through the ice and firn where it is
measured by the station. The reflection has been observed to preserve
polarization~\cite{AriannaIcePaper}. The station is triggered
externally in order to veto radio noise produced by the PCD pulser.



\begin{figure}
  \centering
  \begin{minipage}{0.45\linewidth}
      \centering
        \includegraphics[width=0.9\linewidth]{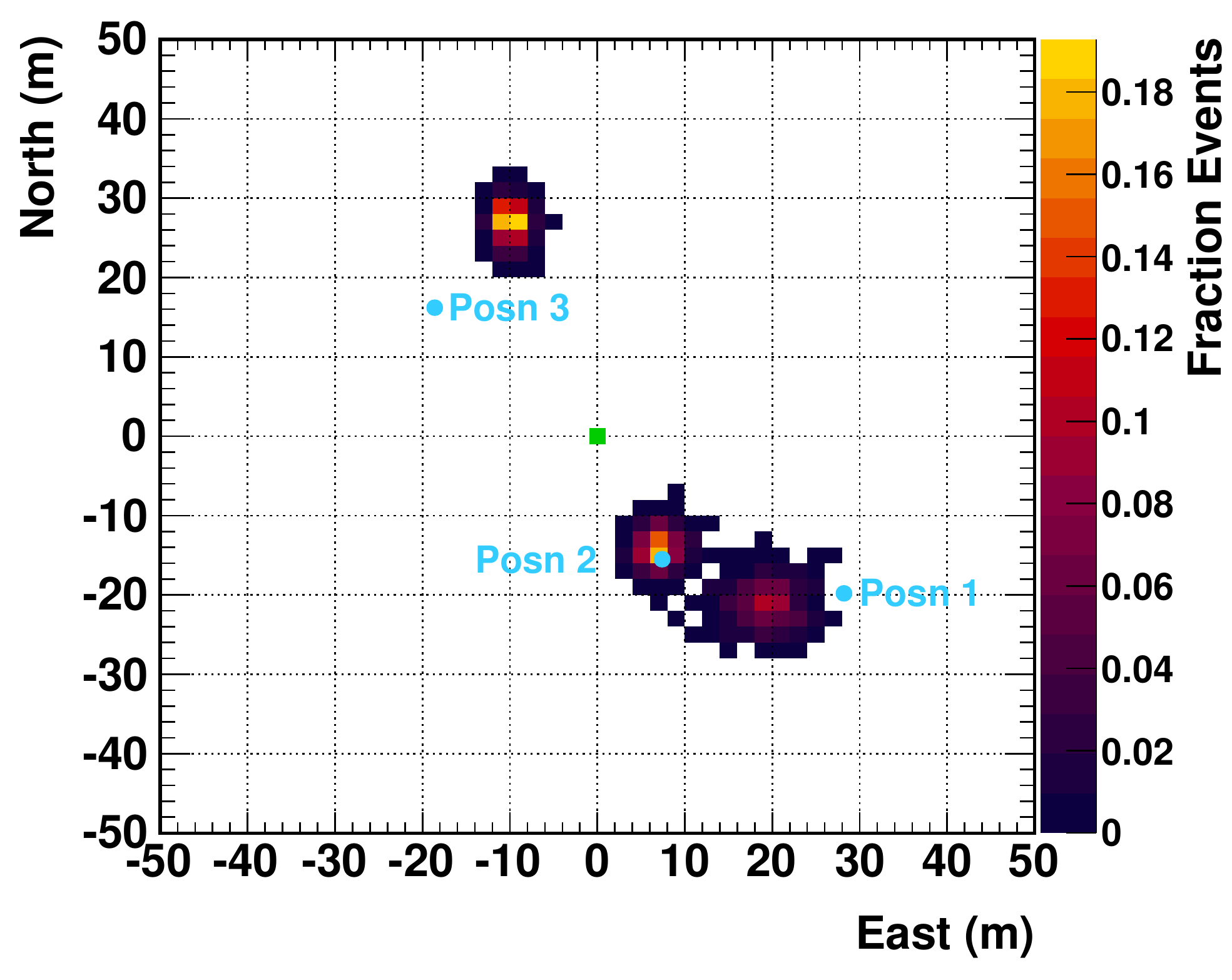}
      \caption{\label{reco:fig:bouncexy}
        The reconstructed calibration pulse transmitter location. The
        true transmitter locations are shown by the blue markers. The
        station sits at the origin, marked by the green square. The
        transmitter LPDA is always oriented such that the plane of its
        tines are offset by 45 degrees from those of each receiving
        LPDA.}
   \end{minipage}
  \hfill
  \begin{minipage}{0.45\linewidth}
    \centering
     \includegraphics[width=0.9\linewidth]{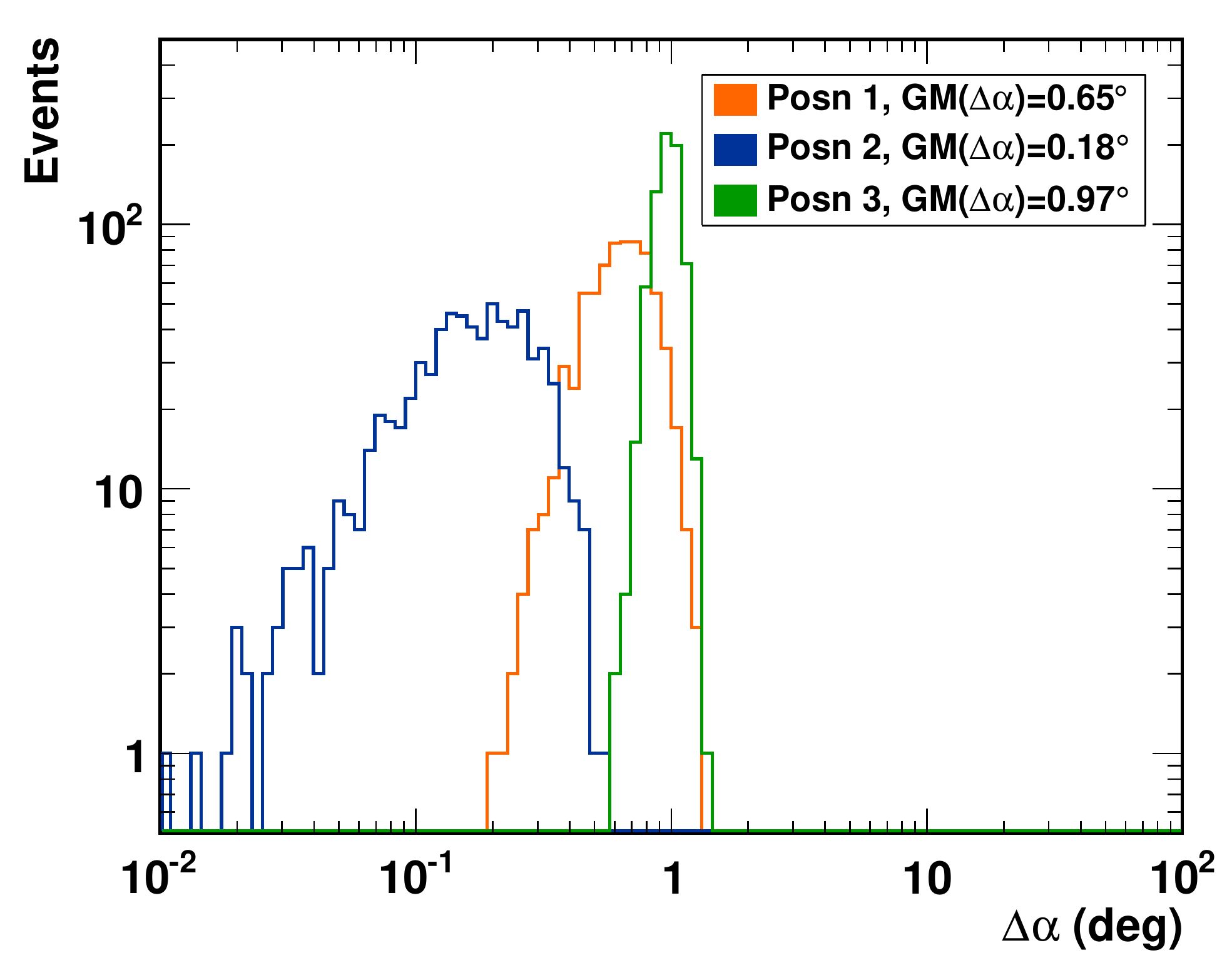}
   \caption{\label{reco:fig:angres} 
     The angular deviation of the reconstructed radio pulse arrival
     direction from the expected direction for events shown in
     \protect\fig{reco:fig:bouncexy}. The resolution is quantified by
     the geometric mean of each distribution, shown in the
     legend. Note that this does not represent a neutrino direction
     resolution.}
   \end{minipage}
\end{figure}

\Fig{reco:fig:bouncexy} shows the reconstructed transmitter location,
relative to the station located at the origin, for events in three
sets of calibration data. The true transmitter positions are shown by
the blue markers. At each location, the transmitting LPDA is oriented
such that its tines are at a 45{\dg} angle to the tines of all four
receiving LPDAs.

The angular deviation between the reconstructed signal direction and
the true direction is shown in \fig{reco:fig:angres}. This represents
the \emph{radio pulse} direction resolution, not a neutrino direction
resolution. The neutrino direction resolution is expected to be on the
order of a few degrees, mainly due to the finite resolution with which
the polarization of the radio signal can be
determined~\cite{Barwick:2014pca}.

\section{Search for Neutrinos in Site B Data}
\label{ana}

The station at Site~B has a configuration that, among currently
deployed {\hra} stations, contains the most pieces of hardware planned
for use in a full {\ari} station. This includes a new signal
digitizing chip, amplifiers with a flatter response over the frequency
bandwidth and a reliable battery. A preliminary search for neutrino
signals in the data taken by this station has been conducted (although
no such signals are expected given the small aperture and short
exposure time of this single station).

Data taken by Site~B from December 10, 2014 (UTC) until the end of the
season, around April 9, 2015 (UTC), is used in the analysis described
in this proceeding. Data collected by the station but not yet
transferred off Antarctica, most of which was taken during the last
weeks of March, is not yet included in the analysis. Also excluded
from the analysis is calibration data and data collected while the
station's communication peripherals were powered on. This latter data
is not particularly noisy and may be included in future analyses, but
has been excluded here for simplicity. The analyzed data set comprises
93.1~days of livetime, corresponding to about a 90\%
uptime~\cite{annaProc} during the data taking periods used in the
analysis. The 10\% downtime is due to the frequency with which the
station transmits its data off Antarctica. This is tuneable remotely
and my be reduced in the future, but was deemed an acceptable cost
given the benefits of receiving station data in near real time.

A set of neutrino simulation events have been generated according to
the procedure described in Ref.~\cite{Barwick:2014pca}. The same
trigger is applied in the generation of the neutrino signal simulation
as has been applied in the Site~B data set. Namely, that two of the
four receiving LPDAs show a bipolar pulse having both the high and low
crests extending beyond the high and low thresholds, each set at four
times the thermal noise RMS.

\subsection{Rejecting Single Frequency Resonance Events}
\label{ana:nhm}

\begin{figure}
   \begin{center}
     \includegraphics[width=0.7\linewidth]{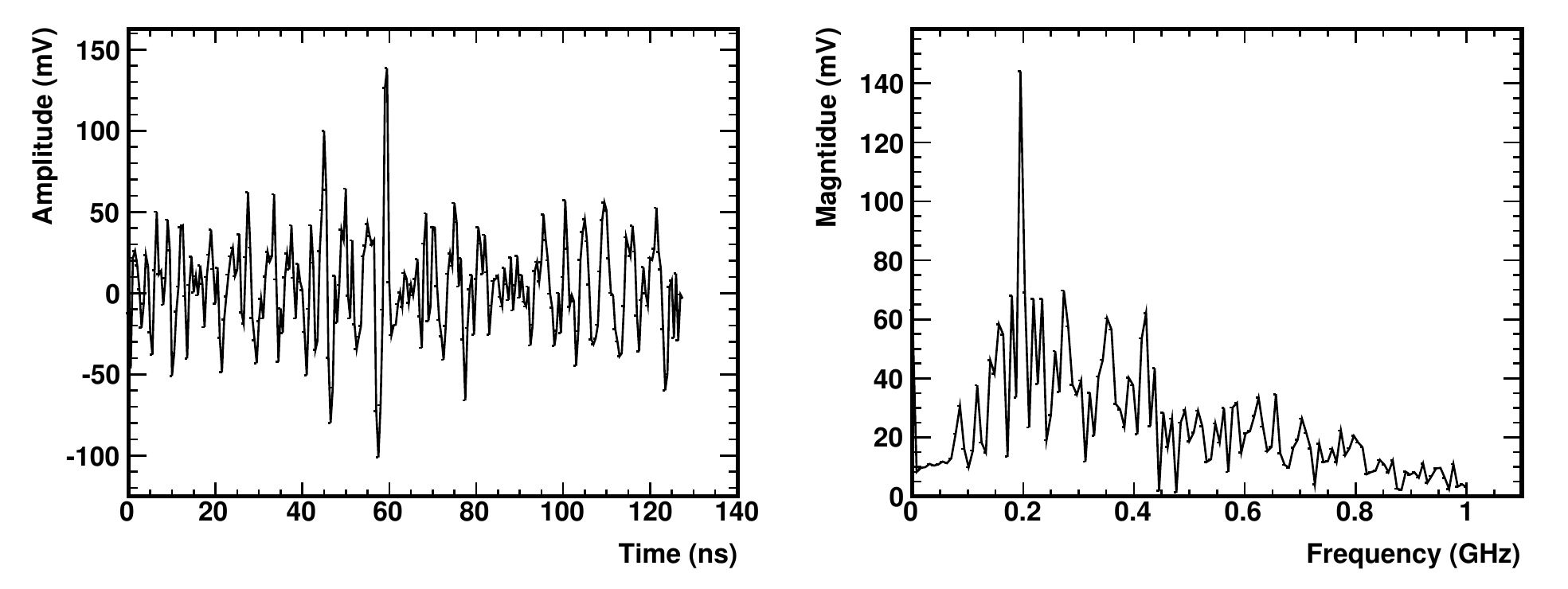}
   \end{center}
   \caption{\label{ana:fig:nhmwvfm}
     Example of a waveform displaying a strong resonance near
     200~{\mhz} that is filtered out by the ${\eta}$ cut. This
     waveform is from the north LPDA. The event was recorded at Site~B
     on March 13, 2015 at 16:46:19 UTC. Left: the waveform in the time
     domain. Right: the waveform in the frequency domain.}
\end{figure}

Some events collected by the station contain a large amount of their
power at a single frequency. Such waveforms stand in stark contrast to
the waveform expected for a neutrino, which contains significant power
across the 50~{\mhz} to 1~{\ghz} frequency band due to the very short
time duration of the Askaryan pulse. These ``sinusoidal'' events in
the data are likely caused by detector electronics, such as radio
noise emitted by the external charge controller of the battery
switching on and off. An example of such an event is shown in
\fig{ana:fig:nhmwvfm}.

Potential neutrino candidate events are required to have a significant
amount of power at several different frequencies. This is done by
first calculating the discrete Fourier transform of the voltage versus
time measured by an LPDA (after amplification). The frequency bin
containing the largest amount of power is then identified; $p_{max}$
represents the amount of power in this bin. The number of frequency
bins that contain more than $p_{max}/4$ on the LPDA is called
${\eta}_{LPDA}$. The smallest ${\eta}_{LPDA}$ value in the event is
defined as ${\eta}$ for the event.

Potential neutrino candidates are required to have ${\eta}>3$
frequency bins.  After applying this cut, 74.3\% of the Site~B data
remains, while 99.3\% of simulated neutrino events that trigger the
station survive.

\subsection{Finding External Pulses}
\label{ana:lik}

The vast majority of triggers recorded by the station are caused by
continuous thermal radio emission. These events are distinguished from
neutrino pulses as they show little to no correlation between parallel
LPDA measurements.

The angular reconstruction procedure described in \sect{reco} is used
to identify events that may have been produced by an external radio
pulse. The angular direction fit likelihood, $-\log(L)$ is required to
be reasonably good for potential neutrino candidate events. The cut on
the likelihood is fairly loose, $-\log(L)<1.7$, for low amplitude
events and becomes stronger, $-\log(L)<0.5$, for high amplitude
events.

\begin{figure}
   \begin{center}
     \includegraphics[width=0.99\linewidth]{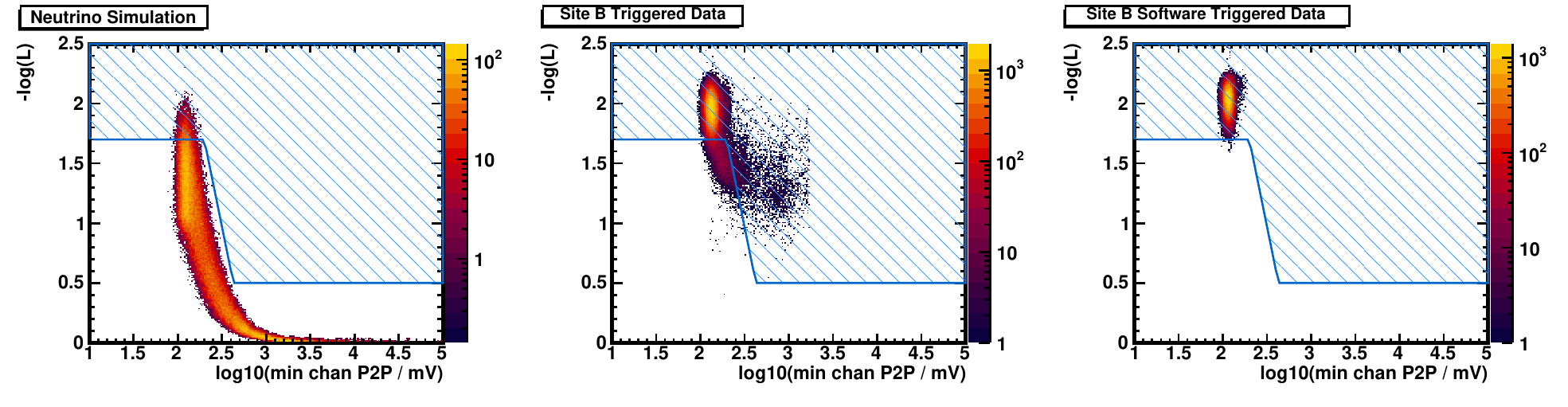}
   \end{center}
   \caption{\label{ana:fig:llvsp2p}
     The fit likelihood versus the log of the peak to peak signal size
     on the LPDA channel with the smallest peak to peak signal in
     three different data sets. Left: the neutrino simulation. Middle:
     Site~B triggered data. Right: Site~B software triggered data
     (pure thermal noise). Events in the blue shaded region do not pass
     the fit likelihood cut. Only events passing the ${\eta}>3$ cut are
     shown for each data set.}
\end{figure}

The fit likelihood cut is shown as a function of the peak to peak
amplitude in \fig{ana:fig:llvsp2p} for neutrino simulations (left),
Site~B triggered data (middle) and pure thermal noise data recorded at
Site~B using software forced triggers (right). The peak to peak
amplitude of an event is quantified by calculating the peak to peak
value of the waveforms on all LPDA channels of an event, and keeping
the smallest peak to peak value. Events outside (below) the blue shaded
region are kept as potential neutrino candidates.

This cut removes nearly all of the purely thermal events recorded by
Site~B. Only 2.7\% of the Site~B data events have both ${\eta}>3$ and
$-\log(L)<f(p)$, where $f(p)$ is a function of the minimum peak to
peak amplitude of the event. The function $f(p)$ is shown by the lower
edge of the blue region in \fig{ana:fig:llvsp2p}. Of the simulated
neutrino events that trigger the station, 96.0\% survive the
application of both cuts.

\subsection{Finding Neutrino-Like Signals}
\label{ana:xi}

After the application of the ${\eta}$ and $-\log(L)$ cuts, many of the
events remaining in the Site~B data set are non-thermal radio pulses.
Much of this data was recorded during periods of high winds as weather
systems pass over the {\ari} site on the Ross Ice Shelf. During the
2014-2015 data taking season, the largest storm occurred during
February. From February 6-8, winds above 40~knots were recorded at New
Zealand's Scott Base, about 100~{\km} from the {\ari} site and over
Minna Bluff. Wind data is not currently available for the {\ari} site
itself. A second important source of the non-thermal Site~B data is
the battery electronics, which emit radio frequency noise when
switching on and off. This occurs toward the end of the season, as the
station alternates drawing power from the battery and solar panels
while periods of strong sunlight become shorter and less frequent.

Simulations of the time dependent waveforms produced by neutrino
signals are used to distinguish background radio pulses from potential
neutrino observations. The time dependent neutrino simulations are
described in Refs.~\cite{Barwick:2014pca, AriannaTemplatePaper}. The
simulations include the measured responses of the {\ari} LPDA as well
as the amplifiers used at Site~B, which play a large role in
determining the expected waveform recorded upon observation of a
neutrino-induced Askaryan pulse.

For a given event, the reconstructed signal direction is used to
determine the arrival direction of the signal in the local frame of
the LPDA. The neutrino waveform expected for a signal probing this
part of the antenna response (and passing through the amplifiers of
Site~B) is then looked up from a table of such neutrino waveform
templates. The template lookup table is binned in
$10{\dg}\times10{\dg}$ wide bins in the local frame of the
antenna. Thus, the reconstruction presented in \sect{reco} is much
more accurate than necessary for the analysis presented in this
proceeding.

The cross-correlation between the chosen neutrino waveform template
and the actual recorded waveform is then calculated. The maximal value
of this correlation (at any time offset) is defined as
$\xi_{LPDA}$. Because the absolute polarization of the potential
neutrino signal is not determined in the data, it is not known whether
the neutrino template or its inverse should compare more favorably to
the recorded waveform. Therefore, both are tried and the larger value
is assigned to $\xi_{LPDA}$.

\begin{figure}
   \begin{center}
     \includegraphics[width=0.75\linewidth]{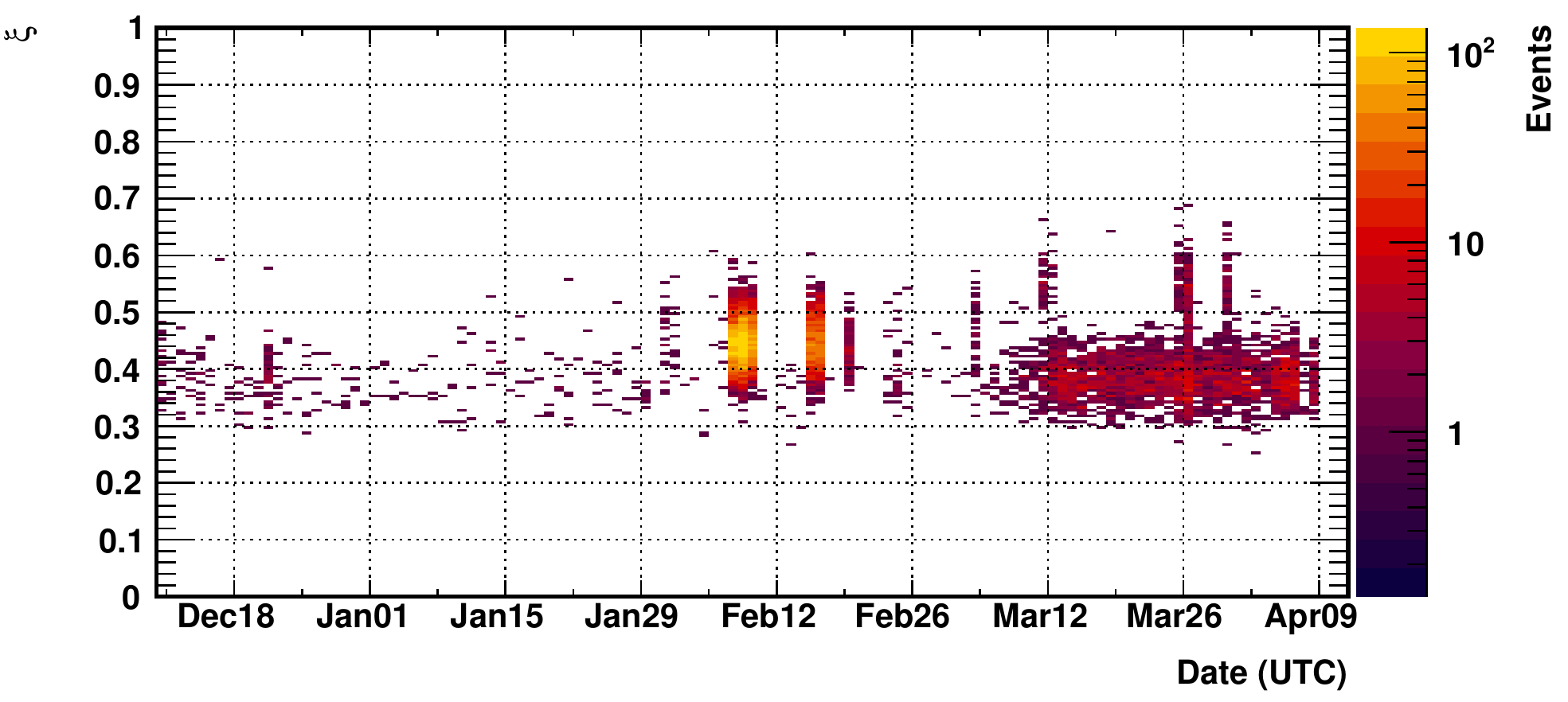}
   \end{center}
   \caption{\label{ana:fig:xiVsT}
     The daily $\xi$ distribution of events recorded by Site~B that
     pass both the ${\eta}<3$ and $-\log(L)<f(p)$ cuts. The solar
     burst is visible in mid December. The large storm in early
     February is also visible. The increased number of events with
     poor $\xi$ values toward the end of the season is a result
     of the higher trigger rates, due to a combination of colder
     temperatures and radio noise emitted by the battery turning
     on and off.}
\end{figure}

A new cut variable, $\xi$, is then defined as the largest $\xi_{LPDA}$
observed on any LPDA channel. \Fig{ana:fig:xiVsT} shows the
distribution of the $\xi$ variable in 24-hour wide bins over the
course of the season for events recorded at Site~B that pass both the
${\eta}<3$ and $-\log(L)<f(p)$ cuts.

Potential neutrino candidate events are required to have
$\xi>0.7$. None of the Site~B data survives the application of all
three cuts, while 85.4\% of the simulated neutrino events that trigger
the station survive the ${\eta}>3$, $-\log(L)<f(p)$ and $\xi>0.7$
cuts. This signal efficiency is comparable to that seen in the
analyses of previous {\hra} season data sets.

The preliminary analysis of the data collected at Site~B between
December 10, 2014 and April 9, 2015 has produced a set of simple
selection criteria that are able to separate potential neutrino
candidate events from radio backgrounds in the data. Data taken from
the other {\hra} stations is currently under analysis. A search for
neutrino signals in the combined 2014-2015 season {\hra} data is
expected to improve upon the diffuse neutrino flux limit presented in
Ref.~\cite{Barwick:2014pca}, assuming no neutrino candidates are found
in the data.

\section{Conclusions}
\label{conc}

The complete {\hra} detector has been installed at the {\ari} site on
the Ross Ice Shelf of Antarctica. All eight stations ran reliably and
took data from the time the installation crew departed the site in
early December until the sunlight faded in late March to early
April. The installation performed during the 2014-2015 austral summer
made use of upgraded detector hardware, some of which is described in
Ref.~\cite{annaProc}.

A radio signal direction reconstruction package has been developed and
applied to both calibration pulse data as well as to triggered data
recorded by the stations. The angular resolution of the reconstruction
is found to be on the order of or better than a degree in both
simulations and calibration data.

The reconstruction has been used in a preliminary search for neutrino
signals in the 93.1~days of data recorded by the station situated at
Site~B. Three cuts are employed by the potential neutrino candidate
search to find events that (a)~have power at multiple frequencies,
(b)~fit well to an incoming plane wave and (c)~have waveforms
resembling those expected for a neutrino.

The vast majority of data removed by these cuts are due to simple
thermal radio noise. Similar to previous seasons, the majority of
non-thermal backgrounds are recorded during periods of high winds at
the {\ari} site. This season, some noise has been observed originating
from the station battery while it is switched on and off by its charge
controller. In future seasons, batteries will be deployed inside the
radio-tight DAQ boxes to eliminate this radio noise source.

No events recorded by the station at Site~B pass the application of
all cuts, while 85.4\% of simulated neutrino events that trigger the
station survive the cuts. This signal efficiency is comparable to that
obtained in analyses of previous {\hra} data taking
seasons~\cite{Barwick:2014pca}. A similar analysis of the data
collected by all {\hra} stations is currently underway, after which a
limit on the diffuse neutrino flux will be placed (assuming no
neutrino candidates are found in the data).


The authors thank the staff of Antarctic Support Contractors,
Lockheed, and the entire crew at McMurdo Station for excellent
logistical support. This work was supported by funding from the Office
of Polar Programs and Physics Division of the US National Science
Foundations, grant awards ANT-08339133, NSF-0970175, and NSF-1126672.
A. Nelles is supported by a research fellowship of the German Research
Foundation (DFG), grant NE 2031/1-1.

\bibliographystyle{JHEP}
\bibliography{BIB}

\end{document}